# SECURING VEHICLE COMMUNICATION SYSTEMS BY THE KLJN KEY EXCHANGE PROTOCOL

Y. SAEZ [1], X. CAO [2], L.B. KISH [1], G. PESTI [3]

[1] *Department of Electrical Engineering, Texas A&M University, College Station, TX*

*77843-3128, USA*

yessica.saez@neo.tamu.edu; laszlo.kish@ece.tamu.edu

[2] *College of Automotive Engineering, Jilin University, Changchun, Jilin*

*130025, China*

caoxiaolin@jlu.edu.cn

[3] *Texas A&M Transportation Institute, Texas A&M University, College Station, TX 77843-3135, USA*

g-pesti@tamu.edu

We review the security requirements for vehicular communication networks and provide a critical assessment of some typical communication security solutions. We also propose a novel unconditionally secure vehicular communication architecture that utilizes the Kirchhoff-law–Johnson-noise (KLJN) key distribution scheme.

## 1. Introduction

During the last years, vehicular communication networks have become an emerging research topic. The main motivation for the deployment of a more intelligent vehicular system is the need to enhance transportation safety and efficiency. In this type of network, vehicles will be equipped with advanced sensing and computing capabilities where communication protocols will enable them to share information with each other and roadside infrastructure. The incorporation of this new range of technology will create a smart net-

*Securing Vehicle Communication Systems by the KLJN key exchange protocol*

work where every vehicle is aware of its surrounding environment. In fact, a great number of applications are under development to improve traffic safety and mobility, and perform financial transactions (e.g. toll collection). These new features will, at some level, improve the quality of life of people and will help to alleviate environmental issues such as pollution and the waste of non-renewable fossil energy [1].

*1.1. Vehicular communication network and nodes*

Figure 1 shows a commonly used vehicular communication architecture model [2–6].

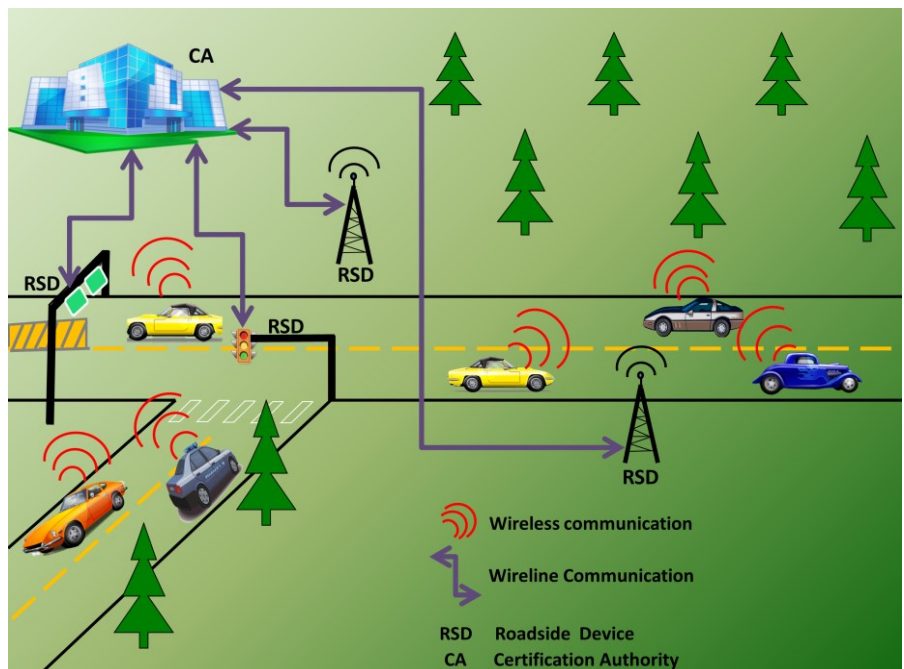

Fig. 1. Vehicular communication network architecture.

Three types of nodes are encountered in this kind of network: Vehicles, Roadside Devices (RSDs) and Certification Authorities (CAs). The vehicles (private or public) and the RSDs are equipped with wireless devices, computing and sensing platforms, and protocol units that will enable them to transmit and receive information. They may be housed in the controller units at signalized intersections, mounted on sensor poles or dynamic message gantries or installed anywhere along the roadside. These devices act as intermediate nodes to vehicles that want to communicate with other vehicles (multiple hops) or with CAs. The certification authority represents a trusted entity in charge of storing and managing information related to the vehicles. Each certification authority is in charge of a specific region and manages all nodes registered with it.

There are also three types of communications taking place in this network: Vehicle-to-Vehicle (V2V), Vehicle-to-Roadside-Device (V2RSD) or as typically called Vehicle-to-Infrastructure (V2I) and Vehicle-to-Certification-Authority (V2CA). The V2V and V2RSD communications use wireless technology, typically the IEEE 802.11p [7], which is an adjustment made to the IEEE 802.11 standard and it has been integrated in the 5GHz Dedicated Short Range Communication (DSRC) [7–8] to add Wireless Access in a Vehicular Environment (WAVE) [8–9]. V2V and V2RSD communications commonly include frequent safety-related messages (warnings) to give the drivers the necessary time to prevent and detect dangerous situations. The V2CA communication requires both wireless and wireline technology, where the RSD links to a wired network connecting the vehicles to the CA. V2CA communication normally includes messages requesting new keys and/or signatures to establish a secure communication with other vehicles or RSDs. Figure 2 illustrates some communications scenarios.

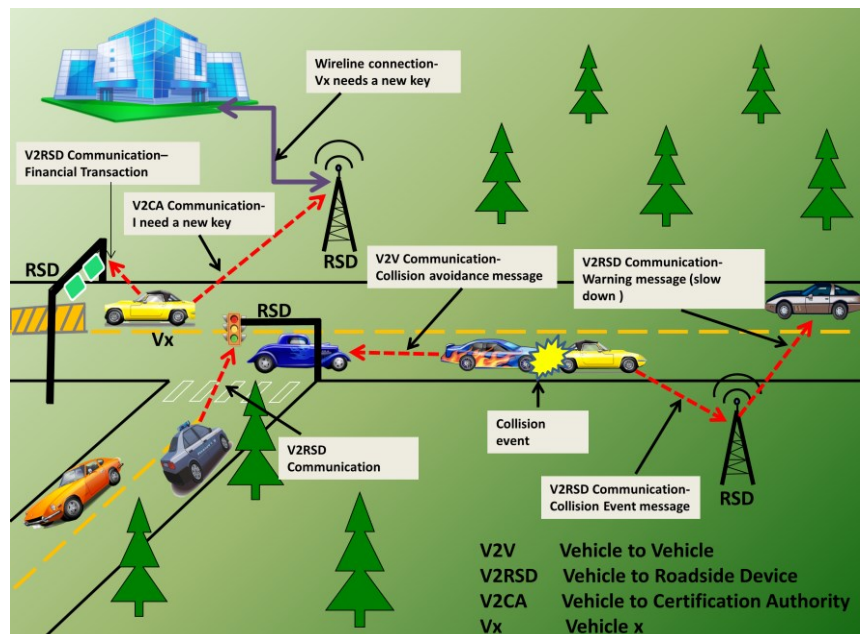

Fig. 2. Communication in vehicular networks.

## 1.2. Vehicular Communication System Security

Even though the integration of new technology and the levels of interconnectivity make the vehicular communication network a more reliable and efficient system, it might also create new vulnerabilities that adversaries could exploit. Since a vehicular network is widely dispersed, its communication infrastructure represents a potential target for malicious users. For instance, an attacker could disseminate false information that could affect the decisions of other drivers. Such attacks could lead to disastrous events such as fatal accidents. Also, a malicious user could monitor the position and/or trajectory of a specific vehicle or listen to financial transactions to steal personal and/or credit card information. Therefore, the safe and successful operation of a vehicular communication network requires the design of very robust security architecture that ensures the protection of private user information without affecting the correct operation of the entire system.

A vehicular communication system should satisfy the following security requirements [4, 10−11]:

- *Authentication*. The receiver must have the capability of validating the sender of a message and must read information only from authenticated senders.
- *Accountability and non-repudiation*. Nodes cannot deny having sent a message. This is crucial when mapping security related events to system entities.
- *Data confidentiality and integrity*. The communication content must remain private and protected the entire time. Unauthorized observers should not be able to read, modify, delete, insert or reorder messages.
- *Availability*. Even a robust network can suffer from attacks and/or incidents that can bring down the communication. Thus, the availability of the system must be supported by alternative means (e.g. network redundancy).

In order to meet the aforementioned requirements, several security methods have been proposed to be implemented in the vehicular networks. Most of them rely on existing security techniques such as symmetric encryption, public key infrastructure (PKI), and digital signature and identity verification, among others. The PKI and the digital signatures are the two most popular methods that have been suggested to be adapted into the vehicular networks. In [12], the authors proposed several security mechanisms by implementing combined signatures and dynamic group creation to protect user's privacy. In [13], the Temporary Anonymous Certified Keys (TACKs) scheme was presented. This key management prevents eavesdroppers from linking keys with a vehicle and a location, without increasing the V2V communication overhead. The authors of [14] proposed a security method by constructing a short range group signature scheme with length under 200 bytes. In [15], a different group signature-based scheme was presented, which achieves the most important security requirements at the same time. In [16], a short, one-time use long chain of keys was proposed. This scheme relies on a combination of traditional digital signatures approaches and light weight broadcast authentication process.

The remainder of this paper is outlined as follows. In Sec. 2, we provide a critical assessment of the security in vehicular communication networks. Section 3 introduces a novel unconditional secure key solution for vehicular communication. Section 4 presents some practical considerations related to the proposed solution. Section 5 concludes the paper.

## 2. Security Considerations

Motivated by the interesting protocols and techniques mentioned above, we discuss some additional concerns related to security in vehicular communication networks.

### *2.1. Secure Short-distance communication*

Even though DSRC is considered a very promising wireless technology for vehicular systems, its effective distance (300~1000 meters [8]) could not be short enough to prevent eavesdroppers from listening to vehicular messages. The V2V and V2RSD communications should be carried out so that other nearby vehicles (parallel or crossing traffic) cannot overhear and/or record the transmitted information. Otherwise, sensitive information (e.g. driver's personal information) could be extracted, which is considered an anonymity/privacy violation. In fact, the recorded information could even be later used by malicious users to create new attack scenarios. Therefore, appropriate security techniques need to be implemented in the wireless technology to be used in V2V and V2RSD communications.

## 2.2. Secure Positioning

Positioning plays an important role in vehicular communication systems. Location information is used in applications like warning messaging, traffic flow monitoring, geographical routing (Geocast), and location-based services [17–18]. In vehicular environments, the time and location of vehicles is usually provided by the Global Positioning System (GPS). This is because currently most vehicles come with a navigation system that includes a GPS receiver which can be used by the vehicular communication system with a little additional cost. However, GPS is not sufficient. Its precision depends on line-of-sight communication with satellites and urban surroundings (e.g. tall buildings and tunnels) degrade its performance. This makes the GPS vulnerable to jamming, spoofing and other kind of attacks from malicious attackers [19].

Secure positioning in vehicular communication networks represents a very challenging area that still needs further investigation. Research must account for techniques that allow vehicles to securely obtain their own and other vehicles' location from the GPS satellites and prevent them from falsifying their position.

## 2.3. Short connection time security

It is very interesting to point out that vehicular communication networks are very dynamic. The constant movement and high speeds of vehicles affect access to the wireless network. The available communication time that vehicles have to exchange messages with each other or with RSDs is very limited. Therefore, short connection time and fast handover methods need to be taken into account in order to have reliable V2V and V2RSD communications. Also, note that traditional security techniques (such as key management) may not work properly. Thus, either the existing security methods must be adapted or new solutions must be designed to fulfill the fast and short connection time requirements.

## 2.4. Security Overhead

We have already pointed out that existing security methods (i.e. PKI, digital signatures) are considered to be the most viable mechanisms to meet the specific characteristics and strict security requirements of vehicular communication networks. Nonetheless, a major disadvantage of these techniques is that they require the broadcasting of many authentication messages that might cause high computational (processing) and/or communication overhead. The processing overhead results from the process of exchanging and verifying digital signatures and certificates, while the communication overhead is a consequence of the extra bits needed for the header and footer security related-fields in the messages [20–21].

Since the overhead increases when the number of vehicles sharing information increases, in high density scenarios the security-related overhead evidently becomes noteworthy, thus affecting the performance of the network. Therefore, it is very important to investigate how to reduce the overhead in the actual solutions and/or propose new techniques and protocols that consider the security-related overhead in its design.

## 2.5. Unconditionally secure techniques

Another topic that is worth considering is that most of the existing security mechanisms for vehicular communications use a software–based key and signature generation and distribution. This means that their performance is based on the assumption that eavesdroppers trying to gain access to security-related information possess limited computational power. Strictly speaking, these techniques offer only computationally conditional security [22]. Therefore, if eavesdroppers can increase their computational power, the keys and digital signatures might be extracted. This would allow them to intercept all the communication between the transmitter and receiver.

In the next section, we propose an unconditionally secure vehicular communication architecture where the information about the key extracted by eavesdroppers is not determined by their computational power but by the laws of physics and the conditions under which the protocols are operating.

### *2.6. Secure V2CA communication*

CAs manage and store very important information associated to vehicles and RSDs, such as location information tables, node identities, and credentials. Before initiating the information exchange with another vehicle or with a RSD, a vehicle needs to obtain security-related information (e.g. certificates) in order to be considered authentic. In this case, the vehicle first communicates with the RSD which then links the vehicle to the CA by using a wireline connection. If this wireline communication is intercepted on the way to/from the CA, important information could be given away. Thus, securing both the V2RSD and RSD2CA communication channels is necessary.

Though there is plenty of research on securing the V2RSD communication, very little attention has been devoted to secure the wireline RSD2CA communication. In the next section, we explain in detail how we can secure the information exchange between the RSD and CA.

### 3. Unconditionally Secure key for Vehicular Communication Networks

In this section, we propose the use of the Kirchhoff-Law-Johnson-Noise (KLJN) scheme to enhance the key exchange in vehicular communication systems. We focus primarily on securing the key distribution of both the wireline and wireless part of the vehicular communication system, seeking to satisfy the security considerations we outlined earlier in this article.

### *3.1. Unconditionally secure key exchange*

Unconditionally secure key exchange, also referred as information theoretic secure key exchange, is considered to be the strictest security condition for key generation/distribution schemes. This is because the security measures in these schemes are determined by information theory, under the assumption that a third party maliciously eavesdropping possesses unlimited resources (i.e. unlimited computational power) to extract information [22]. There are two levels of information theoretic security measures. It can be perfect, which means that the information extracted by the eavesdropper is zero. Another way of interpreting perfect security is that while the eavesdropper is limited only by the laws of physics, the two communication parties can approach perfect security limit provided they have enough resources (e.g. economy, time). Information theoretic security can also be imperfect if there is just a small amount of information leak towards the eavesdropper [23]. It is important to mention that perfectly secure distribution of a key of finite length is not reachable in practice. However, the goal is to come up with schemes that can approach (though never reach) perfect security.

In vehicular communication systems, where security has taken an increasingly important role, there is a need for a new key exchange scheme that can approach a perfect security level. The KLJN-secure key distribution scheme is an unconditionally secure key distribution scheme that is based on Kirchhoff's loop law of quasi-static electrodynamics and the fluctuation-dissipation theorem of statistical physics [23–26]. The security of this scheme is a consequence of the Second law of thermodynamics and its level remains the same even when the eavesdropper's computational power, measurement speed and accuracy are considered to be (hypothetically) infinite. An abstract view of the unconditionally secure KLJN key exchange scheme is shown in Fig. 3 [27–28].

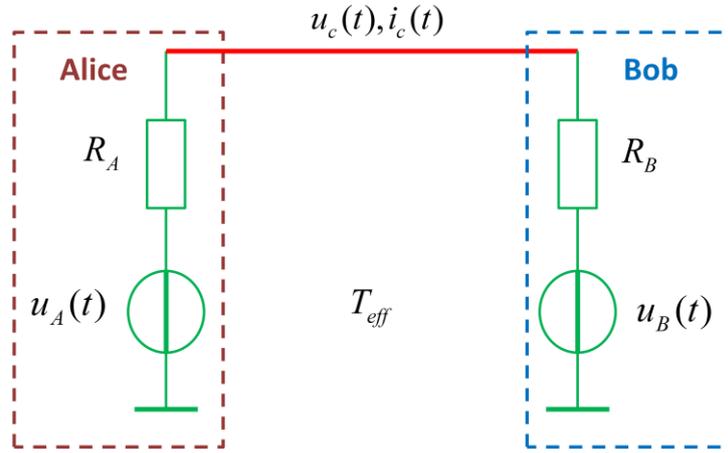

Fig. 3. Illustration of the core Kirchhoff-Law-Johnson-Noise (KLJN) secure key distribution system.

The idealized KLJN scheme can be seen as a wire line connecting two communicating parties, denoted "Alice" and "Bob". At the beginning of each bit exchange period, Alice and Bob—who have identical pairs of resistors $\{R_0, R_1\}$, with $R_1 \neq R_0$ and $R_0$ representing the low (0) bit and $R_1$ the high (1) bit, respectively—randomly select and connect one of these resistors ($R_A$ and $R_B$, respectively) and their thermal-noise-like voltage generators ($u_A(t)$ and $u_B(t)$, respectively) to wire line. Thus, there are four possible ways in which the resistors can be connected to the wire. Alice and Bob can connect the same resistance values to the wires—*i.e.*, the 00- and 11-bit situations. These cases are considered a non-secure bit exchange because an eavesdropper would be able to overhear the communication. The cases when Alice and Bob connect different resistance values—*i.e.*, the 01- and 10-bit situations— represent a secure bit exchange because, as a consequence of the Second Law of Thermodynamics [23–26], an eavesdropper is unable to locate the resistors. Alice and Bob will know that the other party has the inverse of his/her bit, which implies that a secure key exchange takes place.

*Securing Vehicle Communication Systems by the KLJN key exchange protocol*

### 3.2. Network model with unconditionally secure key exchange

Before comprehending the unconditionally secure key exchange protocol for vehicular communication systems, we should first describe our proposed network model. The main goal of this new model is to generate and distribute information theoretically secure keys that are later used to secure information prior to transmission. An abstract view of this vehicular communication architecture, with nodes and authorities, is shown in Fig. 4.

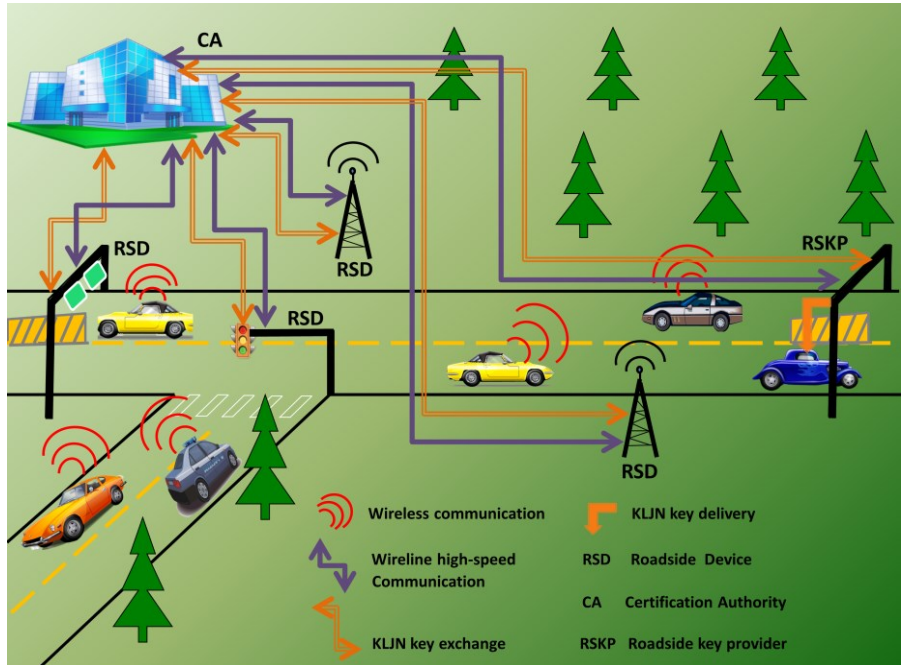

Fig. 4. Vehicular communication network model with unconditional secure key exchange.

As noticed, very few changes have been made to the existing vehicular communication architecture. The network nodes remain the same except for a new node: the roadside key provider (RSKP). The RSKP can be visualized as a gate, and it is in charge of providing the cars with unconditionally secure keys. The communication channel used in this key distribution can be supported by a close proximity communication technology such as Radio Frequency Identification (RFID) [29–31], Near Field Communication (NFC) 31, and/or Near Field magnetic induction communication (see Fig. 5)[32]. Near field magnetic induction communication utilizes an inductive coupling. The operating frequency range is centered on 13.56 MHz on ISO/IEC 18000-3 air interface and offers data transmission rates ranging from 106 kbit/s to 424 kbit/s within a distance of approximately 10 centimeters or less [33].

Since close proximity communication technologies utilize a wireless communication interface, eavesdropping is an important issue [34]. An unauthorized third party could use an antenna to listen the transmitted signals. In order to provide protection against eavesdropping and data modification attacks, a secure channel can be established [34-35]. The authors of [34] proposed a NFC specific key agreement. This key agreement does not require any asymmetric cryptography thus reducing the computational requirements significantly. In [35], a key agreement protocol between a reader and a tag that is resistant in presence of passive adversaries in RFID communication was proposed.

Another change in the network topology is that the RSD2CA communication now utilizes an extra wire for KLJN key exchange. The existing wire line between the RSD and the CA can be kept for high speed communication purposes. Also, an extra wire line between the RSKP and the CA has been included to transmit safety and mobility-related messages.

Table 1 shows a summary of the type of communications between the different nodes in the proposed vehicular communication network with unconditional secure key exchange. It also shows the communication technology utilized and the points at which the KLJN system will be used.

Table 1. Communications in the vehicular network model with unconditional secure key exchange

| Type of Communication | Communication Technology | KLJN system |
|---|---|---|
| V2V | Wireless Communication | No |
| V2RSD and/or RSD2V | Wireless Communication | No |
| V2CA and/or CA2V | Wireless Communication (V2RSD or RSD2V) and Wireline Communication (RSD2CA or CA2RSD) | Yes (wireline segment) |
| CA2RSKP and/or RSKP2CA | Wireline Communication | Yes |
| RSKP2V | Close Proximity Communication | No |

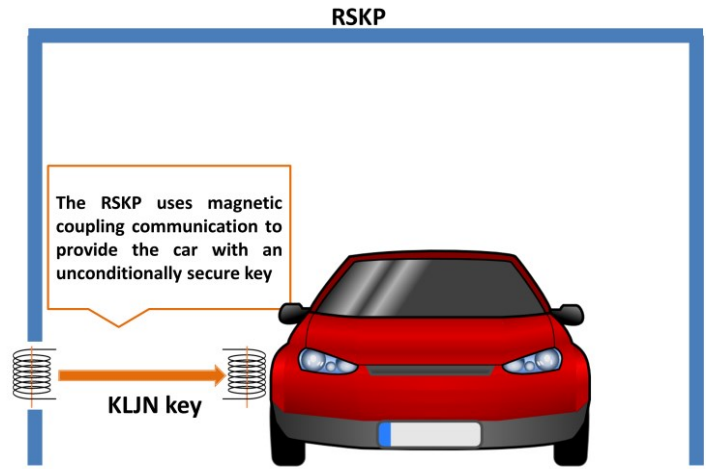

Fig. 5. Abstract illustration of roadside key providers delivering an unconditionally secure key to the vehicle via Near Field magnetic communication. Similar to the transformer principle, the magnetic near-field of two conductor coils is used to couple the initiator device (located at the RSKP) and listening device (located at the vehicle) [33]. Modulation schemes used include: amplitude on/off keying (OOK) with different modulation depth (100 % or 10 %) and Binary phase-shift keying (BPSK)[33].

Under this secure key distribution solution, each node (i.e. vehicle and RSDs) will be assigned a key that does not contain any information related to the identity of a vehicle so user's privacy is preserved. This key will unconditionally secure the information that one node sends to another across the vehicular network. For instance, before a vehicle sends a message, it first signs it with its unconditionally secure key. The receiver of the message has to extract and verify the key of the sender. The protocol used for message authentication and key verification is out of the scope of this paper and will be considered in future works.

## 4. Practical considerations and Future work

### *4.1. KLJN key exchange protocol*

The development of the key exchange protocol to be used in our proposed unconditional secure vehicular communication model is one of the most important subjects to take into consideration in the future. This protocol should comprise a detailed explanation on how keys are generated, distributed and stored. It must also consider the keys' lifetime (duration) and their replacement. Furthermore, the protocol should provide authentication techniques that produce the least possible computational and communication overhead.

### *4.2. Key Length*

The key length is a very important security parameter since it determines the highest security that can be provided. This is because the security of a communication system cannot be better than the security of the key exchange it utilizes. Therefore, a methodology to choose an appropriate KLJN key size to secure vehicular networks is recommended.

### *4.3. Transmission rate*

It is understandable that the transmission rate of distributing the KLJN key is less in comparison to software-based key exchange methods. This is because in this scheme, the duration of the bit exchange period $\tau$ should be long enough to achieve reasonably good noise statistics and securely distinguish between the different resistor situations [24]. Thus, $f_B << B_{KLJN}$ where $f_B \approx 1/\tau$ is the effective bandwidth and $B_{KLJN}$ is the channel noise bandwidth [24, 27–28]. This physical limit determines the tradeoff between the length of KLJN wire, the number of cars served by a single KLJN connection and how well the practical unconditional security will approach the perfect security level. Besides, even though simple and inexpensive ways to improve the speed and security of this key scheme have been proposed [24], a cautious cost-benefit analysis should be carried out to evaluate the cost of additional chip technology and a multi-wire cable.

### *4.4. Technology*

The KLJN key exchange method requires dedicated cables, resistors, close proximity communication technology, statistical tools for bit decision and many other additional technology. Practical implementations of this scheme [25] have shown that this key exchange method is not only low-priced but also extremely robust and almost maintenance-free. However, the amount of KLJN units needed for key distribution will depend on the required key transmission rate and the key length.

## 5. Conclusion

In this Letter, we assessed some concerns regarding the security in vehicular communication networks. Based on this assessment, we outlined how the KLJN system could theoretically be used to achieve unconditionally secure keys to secure vehicular communication networks. The main advantage of this information-theoretic secure key network model is that no computational limitations are placed on the eavesdropper. This means that, with sufficient information about the channel quality and the messages, it is possible to make very accurate statements about the information that is extracted by the eavesdropper.

### Acknowledgements

Y. Saez is grateful to IFARHU/SENACYT for supporting her PhD studies at Texas A&M. X. Cao's contribution is supported by China Scholarship Council.